\documentclass[12pt,twoside]{article}
\usepackage{latexsym,amssymb,amsmath,amsfonts,amscd,amsthm,amstext}
\usepackage[all]{xy}

\swapnumbers
    \theoremstyle{definition}
\newtheorem{def1}{Definition}[section]
\newtheorem{Ax1}[def1]{Axiom}

\newtheorem{cor1}[def1]{Corollary}
\newtheorem{Group}[def1]{Group structure was postulated in 1905}
\newtheorem{Proper-time}[def1]{Proper-time}
\newtheorem{Thomas}[def1]{Thomas rotation}

\newtheorem{groupoid}[def1]{Groupoid relativity}

\newtheorem{conclusion}[def1]{Main conclusion}
        \theoremstyle{plain}

\newtheorem{th1}[def1]{Theorem}

    \theoremstyle{remark}

\numberwithin{equation}{section}

\newcommand{\C}{\mathbb{C}}
\newcommand{\B}{\mathbf{B}}\newcommand{\E}{\mathbf{E}}

\newcommand{\ie}{\textit{i.e.}\,}

\newcommand{\vel}{\mathbf{v}}
\newcommand{\uel}{\mathbf{u}}\newcommand{\xel}{\mathbf{x}}
\newcommand{\wel}{\mathbf{w}}
\newcommand{\zel}{\mathbf{0}}

\newcommand{\bt}{\begin{tabular}{c}}
\newcommand{\et}{\end{tabular}}
\newcommand{\ba}{\begin{array}{c}}
\newcommand{\ea}{\end{array}}

\begin{document}\title{Ternary relative velocity\,\thanks{http://www.worldnpa.org/pdf/ternaryvelocity.pdf}\\{\large Astonishing conflict of the Lorentz group with relativity}}
\author{Zbigniew Oziewicz\;\thanks{A member of Sistema Nacional de Investigadores in {M\'e}xico, Expediente \# 15337.}\,
\thanks{Supported by Programa de Apoyo a Proyectos de Investigac{i\'on} e Innovaci\'on Tecnol\'ogica, UNAM, Grant \# IN104908.}\,\\
Universidad Nacional Aut\'onoma de M\'exico\\Facultad de Estudios Superiores Cuautitl\'an\\C.P. 54714 Cuautitl\'an Izcalli, Apartado Postal \# 25\\Estado de M\'exico\\oziewicz@unam.mx}
\date{
Published in: Vladimir Gladyshev, Editor,\\Physical Interpretations of Relativity Theory\\Moscow 2007, pages 292--303\medskip\\ISBN 978-5-7038-3178-6}\maketitle

\hyphenation{ve-lo-ci-ty ma-ni-fold Min-ko-w-ski Re-la-ti-vi-ty grou-po-id}

\begin{abstract} We are proving that the Lorentz boost entails the relative velocity to be ternary: ternary relative velocity is a velocity of a body with respect to an interior observer \textit{as seen} by a preferred exterior-observer. The Lorentz boosts imply non-associative addition of ternary relative velocities. Within Einstein's special relativity theory, each preferred observer (fixed stars, aether, etc), determine the unique relative velocity among each pair of massive bodies. Therefore, the special relativity founded on Minkowski's axiom, that each pair of reference systems \textit{must} be related by Lorentz isometry, needs a preferred reference system in order to have the unique Einstein's relative velocity among each pair of massive bodies. This choice-dependence of relative velocity violate the Relativity Principle that all reference systems must be equivalent.

This astonishing conflict of the Lorentz relativity group, with the Relativity Principle, can be resolved in two alternative ways. Either, abandon the Relativity Principle in favor of a preferred reference system [de Abreu 2004; de Abreu \& Guerra 2005, 2006]. Or, within the Relativity Principle, replace the Lorentz relativity group by the relativity \textit{groupoid}, with the choice-free binary relative velocities introduced by Minkowski in 1908. Binary - means that the relative velocity is a function of a pair of reference systems. Following Minkowski [1908], we consider the binary relative velocity to be the Minkowski space-like vector (and \textit{not} the Minkowski bivector as it is in the Hestenes theory [Hestenes 1974]).

An axiomatic definition of the kinematical unique binary relative velocity as the choice-free Minkowski space-like vector, leads to the groupoid structure of the set of all deduced relativity transformations (instead of the Lorentz relativity group), with the associative addition of binary relative velocities.

Observer-independence, and the Lorentz-invariance, are distinct concepts. This suggest the possibility of formulating many-body relativistic dynamics without Lorentz/Poincare invariance.\end{abstract}


\noindent\textbf{Physics and Astronomy Classification Scheme (PACS) 1999/2000.}\\ \indent\quad 03.30.+p Special relativity.\smallskip


\pagestyle{myheadings}\markboth{\quad
\emph{Zbigniew Oziewicz}\hfill}{\hfill\emph{Ternary relative velocity}\quad}

\tableofcontents

\section{Binary relative velocity} We formulate the relativity theory in the ancient Aristotelian spirit, that precede the Galileo and Descartes revolution, by postulating that the concept of relative velocity must be the primary concept of the theory. In Aristotelian spirit, time and space are attributes of relative velocity. Our postulate leads to a groupoid category. This formulation could be contrasted with the sacred understanding of the relativity where the primary concept is the Lorentz (Poincar\'e) \textbf{group} of isometries in the Felix Klein spirit: the group must be primary!

The main objective of the present short note is the concept of a ternary relative velocity. A ternary relative velocity is a velocity of a massive body with respect to an interior observer \textit{as seen} by a preferred exterior-observer. It will be convenient to present the ternary relative velocity, in terms of the binary velocity (of a medium), relative to an observer, used by Minkowski in 1908.
Following Minkowski, we identify a reference system (a massive observer) with a normalized time-like Minkowski vector field, $P^2=-1.$

\begin{def1}[Observers and subjects-observed]\label{owners} Each non-zero space-like Minkowski vector, $\wel\neq\zel,$ possess the following pair of two-dimensional sub-manifolds (hyperbolises) of potential observers of $\wel,$ and of massive subjects that are potential possessors of $\wel,$
\begin{gather}O_\wel\equiv\{A^2=-1,A\cdot\wel=0\},\qquad
S_\wel\equiv\{B^2=-1,B\cdot\wel=\wel^2\}.\label{D}\end{gather}\end{def1}
The notation, $A\in O_\wel,$ is equivalent to the phrase that an observer $A$ is observing a massive body, $\sqrt{\wel^2+1}\,A+\wel\in S_\wel,$ possessing a space-like Minkowski vector $\wel.$

Analogously, $B\in S_\wel,$ means that a massive body $B$ possess a space-like Minkowski vector $\wel,$ relative to an observer, $\frac{B-\wel}{\sqrt{\wel^2+1}}\in O_\wel.$

In what follows, for a bounded space-like Minkowski vector, such that, $\vel^2<c^2,$ the Heaviside - FitzGerald - Lorentz scalar factor is denoted by,
\begin{gather}\vel\longmapsto\gamma_\vel\equiv\frac{1}{\sqrt{1-\frac{\vel^2}{c^2}}},\qquad
\wel\equiv\gamma_\vel\frac{\vel}{c}\quad\Longleftrightarrow\quad\frac{\vel}{c} \equiv\frac{\wel}{\sqrt{1+\wel^2}},\\\gamma_\vel=\sqrt{1+\wel^2}.\end{gather}
Sometimes, for simplicity of formulas, the scalar magnitude of the light velocity is set, $c^2=1.$

\begin{Ax1}[Binary relative velocity]\label{Axiom} Let $A\in O_\uel,$ and
$\uel^2<c^2.$ Then, and \textit{only} then, $\exists\,!$ a massive body,
$B=b_\uel A\equiv\gamma_\uel(A+\frac{\uel}{c})\in S_\uel,$ moving with a
velocity $\uel$ \textit{relative} to $A.$ This imply, $\gamma_\uel=-A\cdot B.$
The space-like relative velocity $\uel$ is said to be \textit{binary},
\begin{gather}\frac{\uel}{c}\equiv\frac{\varpi(A,B)}{c}=\frac{B}{-B\cdot A}-A\qquad\Longrightarrow\qquad 1-\left(\frac{\uel}{c}\right)^2=\frac{1}{(A\cdot B )^2}=\frac{1}{\gamma_\uel^2}\label{Ax2}\end{gather}\end{Ax1}

\section{Ternary relative velocity} Minkowski identify a reference system with a time-like vector field, and defined the special relativity by means of the following single axiom.

\begin{def1}[Special relativity]\label{Minkowski} Any two reference systems must be connected by the Lorentz isometry, \ie, by the isometry acting on \textit{all} vectors, including \textbf{not} time-like vectors.\end{def1}

The Minkowski Definition-axiom \ref{Minkowski} does not need explicitly the concept of relative velocity, and leads to the cornerstone of XX century physics: Lorentz group-covariance.

The vector space of the Grassmann bi-vectors inside of Clifford algebra, is the Lie algebra of the Lie group of isometries. Each Minkowski bi-vector, $P\wedge Q,$ generate an isometry $$P\wedge Q\quad\hookrightarrow\quad L_{P\wedge Q}\in O(1,3).$$

\begin{def1}[Ternary relative velocity]\label{ternary} Let $\{P,A,B\}$ be a three-body massive system given by a set of three time-like Minkowski vectors, and let $P\in O_\vel,$ be an actual observer of a bounded space-like velocity $\vel,$ $\vel^2<c^2.$ The Minkowski velocity vector $\vel$ of a Bob $B$, relative to Alice $A,$ as seen/measured by a preferred observed $P,$ is said to be \textit{isometric}, or \textit{ternary}, or \textit{the Einstein velocity}, if it is defined in terms of the isometric Lorentz boost as follows
\begin{gather}\overline{\vel}\equiv\gamma_\vel\vel/c,\qquad
L_{P\wedge\overline{\vel}}\in O(1,3),\qquad \fbox{\quad$L_{P\wedge\overline{\vel}}\;A=B$\quad}\label{boost}\end{gather}\end{def1}

The above definition is motivated by the following theorem.

\begin{th1}[Isometry-link problem (Oziewicz 2005, 2006)]\label{Oz} For the massive three-body system given in terms of the three time-like vectors $\{P,A,B\},$ the Lorentz-boost-link equation for unknown space-like Minkowski vector $\wel,$ $L_{P\wedge\wel}\,A=B,$ has the unique solution, $\overline{\vel}=\wel(P,A,B).$
A ternary relative velocity, parameterizing the Lorentz boost \eqref{boost}, looks like a kind of subtraction of absolute/binary velocities,
\begin{gather}\vel(P,A,B)\;=\;P\cdot(A+B)\,\frac{(P\cdot B)
\varpi(P,B)-(P\cdot A)\varpi(P,A)}{(P\cdot A)^2+(P\cdot B)^2-1-A\cdot B},\label{subtraction}\\
\vel(P,A,B)=-\vel(P,B,A)\quad\simeq\quad i_{gP}\{P\wedge(B-A)\},\label{reciprocal}\\
\gamma_{\text{ternary}}=\frac{(P\cdot A)^2+(P\cdot B)^2-A\cdot B-1}
{2(P\cdot A)(P\cdot B)+A\cdot B+1}\quad\neq\quad-A\cdot B=\gamma_{\text{binary}}.\label{gamma}\end{gather}\end{th1}

\begin{cor1}\label{coplanar}Consider co-planar system of massive bodies, $P\wedge A\wedge B=0.$ In this particular case a ternary relative velocity \eqref{subtraction} is reduced to the expression presented by de Abreu \& Guerra [2005, page 74],
\begin{gather}\left.\vel(P,A,B)\right|_{P\wedge A\wedge B=0}\;
=\;\frac{\varpi(P,B)-\varpi(P,A)}{1-\varpi(P,B)\cdot\varpi(P,A)/c^2}.\end{gather}
\end{cor1}

\begin{th1} The scalar magnitudes of the binary and ternary relative velocities, \eqref{Ax2}-\eqref{subtraction}-\eqref{gamma}, coincide, $\gamma(\text{binary})=\gamma(\text{ternary}),$ if and only if the three-body system is co-planar,
\begin{gather}\gamma_{\vel(P,A,B)}=-A\cdot B\qquad\Longleftrightarrow\qquad P\wedge A\wedge B=0.\end{gather}\end{th1}

The Minkowski Definition-axiom \ref{Minkowski} does not need explicitly the concept of relative velocity, and leads to choice-dependent, $P$-dependent, ternary relative velocity among Alice and Bob. Each reference system $P$ (that could be interpreted as the physical fixed stars, aether, etc), gives the unique Einstein's reciprocal relative velocity among each pair of massive bodies, expression \eqref{subtraction},
\begin{gather}\{A,B\}\quad\xrightarrow{\quad\text{preferred}\; P\quad}\quad\vel(P,A,B).\end{gather}
Definition \ref{ternary} and Theorem \ref{Oz} tells that the Lorentz-boost is \textit{not} unique. It is true that all textbooks of special relativity, define the Lorentz boost as the `unique' basis-dependent matrix, see for example in [Ungar 1988], \begin{gather}\text{Lorentz boost}\quad\equiv
\quad\begin{pmatrix}\gamma&-(v/c)\gamma&0&0\\-(v/c)\gamma&\gamma&0&0\\
0&0&1&0\\0&0&0&1\end{pmatrix}\label{Barut}\end{gather}
When a priori accepting the above matrix definition \eqref{Barut}, it is hard to imagine the existence of the another boost, because the above matrix is fixed by velocity parameter, and there is no option for alternative boost. We like to point that our definition of basis-free $P$-dependent Lorentz boost \eqref{boost}, is reduced to the above `unique-matrix', in the particular basis, when, $P\simeq(1,0,0,0),$ and $\overline{\vel}\simeq(0,-v/c,0,0).$ This explain why someone insists that the `Lorentz boost is unique!'.

Contrary to the matrix-statement, \eqref{Barut}, a Lorentz boost-link from $A$ to $B$ is not unique \eqref{boost}. Lorentz boost is choice-dependent, depends on the choice of the preferred observer $P.$ Equivalently, the Lorentz boost depends on the choice of the non-unique embedding of the rotation group as the non-normal sub-group of the Lorentz group, $O(3)\hookrightarrow O(1,3).$ Each Lorentz boost is given by bi-vector, therefore can not be parameterized by a space-like velocity vector alone. The Minkowski Definition-axiom \ref{Minkowski} means, that when a preferred vector $P$ is \textit{not} chosen, there is a bunch of Lorentz-link transformations from $A$ to $B,$ generated by many different bi-vectors, $$P\wedge\wel\quad\neq\quad P'\wedge\wel'\quad\neq\quad\ldots,$$ but there is \textbf{no} relative velocity among massive bodies. To have exactly one Einstein's velocity of Bob relative to Alice, we need to chose some one and only one reference system to be preferred. This choice-dependence of Einstein's relative velocity violate the relativity principle stating that \textit{all} reference systems must be equivalent. Within the Minkowski (Lorentz-relativity-group)-Definition-axiom \ref{Minkowski}, an equivalence of all reference systems is not possible: the different choices of a preferred system $P$ in Definition \ref{ternary}, lead to distinct Einstein's velocities of Bob relative to Alice. Each observer $P\in O_\wel,$ gives rise to his own isometric Lorentz $P$-boost $L_{P\wedge\wel},$ and therefore the Lorentz boost is not unique.

We distinguish conceptually the \textit{relativity}-group (permuting the reference systems), from the concept of a symmetry-group. The Lorentz group is a symmetry-group of the metric of the empty space-time, it is a group of isometries. The Lorentz group is also a symmetry-\textit{sub}group of conformal symmetry of the Maxwell equations. However, a priori, a set of all relativity transformations among massive reference systems need \textit{not} to be a group, and need not to be a symmetry-group of some other mathematical structure. We remind that, for example, the light-like vectors do not represent reference systems, therefore, a priori, they do \textit{not} need to be in the domain of relativity transformations. The present paper deals exclusively with the Lorentz group as the relativity-group.

In fact, the main problem of relativity theory is \textit{not} about the special relativity identified with the Minkowski axiom \ref{Minkowski}, it is \emph{not} about the interpretation of the Relativity Principle, it is \emph{not} about the clock's synchronization, it is \emph{not} so much about time measurements, it is \emph{not} about one-way \& two-way light velocity. The main problem is the coordinate-free definition of the concept of the relative \textbf{velocity}. Textbooks devote a lot of attention to the important dynamical concepts of acceleration and force, however, following Galileo, the kinematical velocity is defined in coordinate-dependent way, such that the eventual conceptual distinction among absolute and relative velocities, among non-relativistic and relativistic velocities, is obscure in the expression like `$x=vt$'. How such `$v$' depends on the choice of the reference system? how depends on the choice of mathematical coordinates? What are abstract properties of the set of all relative velocities, including the law of addition-composition of relative velocities?

The main question is about the precise axiomatic coordinate-free definition of the concept of relative velocity. The distinct theoretical conceptual definitions of relative velocity must proceeds to the experimental measurements. How to measure, without understanding what concept are you going to measure? Textbooks repeat `a passenger sitting in a moving train is at rest in relation to the train, but it is in a motion in relation to the ground'. How one can see this in a coordinate expression `$x=vt$'? Where in these three symbols, $\{x,t,v\},$ there is a train, passenger, ground? We need to have something like at least two-variable expression
\begin{gather}\vel(S,\text{passenger})=\begin{cases}\zel_{\text{train}}&\text{if $S$ is a train,}\\\neq\zel&\text{if $S$ is a ground,}\end{cases}\qquad
\vel(\text{train},\text{train})\equiv\zel_{\text{train}}.\end{gather}

From this main point of view of the precise definitions of the concept of a relative velocity, Definition \ref{ternary}, is the very precise mathematical definition: the Einstein's ternary relative velocity, expression \eqref{subtraction}, is coordinate-free, and basis-free. From this definition one can deduce \textbf{all} properties of such Einstein's ternary velocities that we are listening in Section \ref{ter} (including their non-associative addition), and one can easily deduce the coordinate expression for any system of coordinates (we left this to interested students). Reader do not need to like the definition of a ternary relative velocity. Each reader could invent his own different definition of the relative velocity. However, I hope that all readers agree that Definition \ref{ternary} is the precise mathematical definition. We do not need at this moment enter to the experimental problems of the measurements of ternary velocities, nor to the question does the Nature like or dislike the mathematical definition of ternary relative velocity?

\begin{conclusion} If we believe (or if we postulate) that:\\(i) There must be one and only one relative velocity among each pair of massive bodies.\\(ii) Each pair of massive bodies must be related by an isometry, Definition \ref{Minkowski}.\\\textit{Then}, a ternary velocity needs the choice of the preferred reference system.

Definition \ref{ternary} say: yes, there is one and only one relative velocity among each pair of massive bodies, \textit{provided} that the preferred reference system (absolute space) was chosen. Therefore, the concept of the preferred reference system, an aether, is built in the Einstein velocities.\end{conclusion}
\begin{itemize}\item Preferred reference system is attractive for explaining the non-symme\-tri\-cal ageing of twins (Herbert Dingle, Rodgrigo de Abreu \& Vasco Guerra, Subhash Kak).\item Preferred reference system is meaning-less for believers in Lorentz-covariance as the corner\textbf{stone} of Physics (the Lorentz relativity-group is Sacred!). \item Preferred reference system is a fault for believers in the Relativity Principle (no need for the choice of a preferred reference system in order to have a relative velocity).\end{itemize}

Nevertheless one can accept special relativity with Lorentz-relativity-group axiom, Definition \ref{Minkowski}, admitting that in this case we must violate the Relativity Principle in order to have the unique Einstein's relative velocity. The uniqueness of relative velocity needs the choice of one reference system to be preferred. Therefore, the special relativity with Lorentz relativity-group, not only is perfectly consistent with a special system of reference (alias preferred system, alias Einstein's lost frame, alias aether, fixed stars, etc), as concluded independently in [de Abreu 2004; de Abreu \& Guerra 2005; Guerra 2006; Kak], but, such special Lorentz-group-relativity does \textbf{not} exist at all, without preferred reference system. No choice of a preferred vector $P$ in Definition \ref{ternary}, no velocity $\vel$ of Bob relative to Alice.

We see the astonishing conflict! The special relativity with Lorentz-relativity-group axiom necessarily contradicts to the Relativity Principle. Not all reference systems can be considered any more as equivalent! Instead, one reference system necessarily must be chosen in order that each pair of massive bodies has the unique relative velocity.

\section{Rodrigo de Abreu: restricted Relativity Principle}
De Abreu proposed to abandon the Relativity Principle in favor of `restricted Relativity Principle' that allows the absolute space with a preferred reference system, referred to as `the Einstein's lost frame'. This idea was future developed in [De Abreu 2002, 2004; De Abreu \& Guerra 2005; Guerra \& de Abreu 2006]. The velocity relative to the preferred reference system is said to be the \textit{absolute} velocity, and a velocity relative to non-preferred system is said to be the Einstein velocity [De Abreu 2004]. The starting point of De Abreu (and jointly with Guerra) is the observation that the Einstein synchronization of clocks can be made in one and only one reference system. Analysis of the clock synchronization (related to one-way versus two-ways light velocity) leads Authors to consider the abandoning of the Relativity Principle (that all reference systems are equivalent).

\begin{Ax1}[De Abreu]\label{Abreu} The one-way light-velocity is source-free in \textbf{one and only one} reference system. This reference system is said to be the Einstein lost-frame.\end{Ax1}

Definition \ref{ternary} of the ternary relative velocity, implies that \textit{all} these velocities are reciprocal, \eqref{reciprocal}. If $\vel$ is the Einstein velocity of Bob relative to Alice, then the Einstein velocity of Alice relative to Bob is $-\vel,$ for \textit{each} choice of the preferred reference system $P.$ However, this is not the case in special relativity theory in [de Abreu \& Guerra 2005]. De Abreu \& Guerra in their monograph introduced three different concepts of velocities. The \textit{absolute} velocity, denoted here by $\vel_{AG},$ is a velocity of reference system, say Bob $S',$ relative to the preferred absolute space at rest (say Aether $S$), for the chosen event $P$ [Abreu \& Guerra 2005, pages 41-42, Figure 3.10], \begin{gather}\vel_{AG}(P,S,S')\equiv\vel_{AG}(\text{event},\text{Aether},\text{Bob})=
\frac{\xel}{t}-\left(\frac{t'}{t}\right)^2\,\frac{\xel'}{t'}.\label{AG}\end{gather} This \textit{absolute velocity} is defined in terms of the coordinates of an event $P,$ therefore, it is the `absolute velocity' of Bob `as seen by an event $P$'. It is not obvious that the choice of a spectator-event $P$ must be irrelevant for Definition \eqref{AG} of absolute velocity. One can guess that `preferred spectator-event $P$' (denoted accidently on Figure 3.10 on page 41 by the first letter of `preferred'), is probably assumed to be in the rest relative to the Aether (or it is in the rest relative to the Bob?). What would be if the event $P$ will be chosen to be neither in the rest relative to Aether, nor in the rest relative to the Bob?

De Abreu and Guerra define the \textit{Einstein speed} that is the reciprocal Einstein velocity among two frames [Abreu \& Guerra 2005, page 74], and this concept coincide with our ternary relative velocity for co-planar systems, Corollary \ref{coplanar}. Moreover, the Authors have also the `Rodrigo' non-reciprocal \textit{relative speed} defined on page 44. What seems to me to be the most essential peculiarity of the Abreu \& Guerra's special relativity theory, that their absolute-velocity $\vel_{AG}$ \eqref{AG}, is \textit{not} reciprocal, and the velocity of the Aether as measured by Bob has much higher scalar-magnitude when compared with the absolute velocity of the Bob as measured by the Aether, [Abreu \& Guerra 2005, page 42], viz. \begin{gather}\vel_{AG}(\text{event},\text{Bob},\text{Aether})=-(\gamma_\vel)^2\,
\vel_{AG}(\text{event},\text{Aether},\text{Bob}).\label{AG1}\end{gather}

\section{Relativity-groupoid as alternative} Does exist some alternative theory that is completely compatible with the Relativity Principle? The alternative philosophy is to keep Relativity Principle, however, \textit{replace} the Lorentz relativity-group (with choice-dependent Einstein's relative velocity), by a relativity groupoid (with choice-free axiomatic binary relative velocity). The alternative is to consider the relative velocity among massive bodies as the primary concept, and then derive/deduce/\textit{define} the transformation among reference systems in terms of this primordial, given a priori, binary relative velocity.

The Einstein special relativity theory, consider the isometry Lorentz relativity transformation, $L\in O(1,3),$ as the primordial concept, and the relative velocity as the derived concept,
\begin{center}$\left.\begin{tabular}{c}group of isometries $\equiv$\\relativity transformations\end{tabular}\right\}\quad\Longrightarrow$\quad ternary relative velocities\end{center}

The possible alternative is to axiomatize the concept of the unique binary kinematical relative velocity as the primordial concept, and \textit{derive} the relativity transformation among reference systems in terms of this given choice-free binary velocity,
\begin{center}binary relative velocities\quad$\Longrightarrow\quad\left\{ \begin{tabular}{c}relativity transformations\\groupoid category \end{tabular}\right.$\end{center}
Then, could we have a hope that such \textit{derived} set of all relativity transformations, parameterized by the choice-free axiomatized relative velocities, will coincide with the \textit{group} of Lorentz isometries (parameterized by the choice-dependent Einstein's velocities)?

The binary relative velocity-morphism is the choice-free, Definition \ref{owners}, and Axiom \ref{Axiom}. We claim that such axiomatic binary velocity can \textit{not} parameterize the isometric Lorentz transformation. The one reason, among other, is that the domain of the action parameterized by the choice-free binary velocity, $\frac{\vel}{c}\equiv\frac{\wel}{\sqrt{1+\wel^2}},$ is restricted to the two-dimensional sub-manifold of Minkowski vector fields, $O_\wel\xrightarrow{\;\wel\;}S_\wel,$ Definition \ref{owners}. The set of all relativity transformations parameterized by binary relative velocities has the structure of a groupoid category (that is not a group), and the addition of binary velocities is associative.

\begin{def1}[Groupoid category] A category is said to be a \textit{groupoid category}, if and only if every morphism has a two-sided inverse. In particular a \textit{group} is a groupoid one-object-category, with just one object, hence with universal unique neutral element-morphism. A groupoid category is said to be
\textit{connected} if there is an arrow joining any two of its objects.\end{def1}

We propose to formulate the physics of relativity in terms of the groupoid category of observers, keeping strictly the most democratic interpretation of the Relativity Principle that \textit{all} reference systems are equivalent. The groupoid relativity starts with the axiomatic definition of the binary relative velocities-morphisms, that are choice-free, Axiom \ref{Axiom}, and conclude that these relative-velocities can not parameterize the isometric Lorentz transformations.

This formulation is perfectly consistent with the principle of relativity, because all reference systems are equivalent, and there is no need for the choice of the preferred reference system. In particular, within the groupoid relativity, the inverse of relative velocity is not reciprocal, however this groupoid non-reciprocity is different from non-reciprocity in Abreu-Guerra theory \eqref{AG1},
\begin{gather}\vel^{-1}\cdot\vel=\begin{cases}-\;\;\gamma_\vel\hspace{2mm}\vel^2,
\hspace{5mm}|\vel^{-1}|=|\vel|&\text{in groupoid relativity,}\\
-(\gamma_\vel)^2\vel^2,\quad|\vel^{-1}|\neq|\vel|&
\text{in Abreu \& Guerra relativity.}\end{cases}\end{gather}
Within the groupoid relativity, the inverse is an involutiv operation, $(\vel^{-1})^{-1}\equiv\vel,$ whereas this is not the case within Abreu \& Guerra theory.

Zaripov consider anisotropic Finsler spacetime, and proved that the Finslerian inverse operation (of relative velocity) is non-reciprocal, $\vel^{-1}\neq-\vel,$ and, in Zaripov's theory, it is also non-involutive, $(\vel^{-1})^{-1}\neq\vel,$ [Zaripov 2006].

The associative $\circ$-addition of binary relative velocities is a trivial corollary that follows from two related concepts: the binary relative velocity is a categorical morphism, and a derived groupoidal boost can not be an isometry, Axiom \ref{Axiom}. In the consequence, a bivector, $(\uel^{-1})\wedge\uel\neq 0,$ does not vanish.

The relativity-groupoid predicts exactly the same time-dilation as the relativity-Lorentz-group, however no material rod contraction.
One can test experimentally the Lorentz-group relativity versus groupoid-relativity within mo\-ving media electrodynamics, and we refer to related proposals by Gladyshev et al. [2000, 2005]. Consider two reference systems given by Minkowski time-like vectors, Alice and Bob, and let $\uel$ be a space-like Minkowski vector of binary velocity of Bob relative to Alice, \ie the relative velocity as measured by Alice, $\uel\cdot A=0.$ Let, $\E'$ and $\B',$ be electric and magnetic fields measured by Bob $=\gamma(A+\uel/c).$ Let, $\E,$ and $\B,$ be the electric and magnetic fields as measured by Alice. Then,
\begin{gather}\uel\times\B=\zel\quad\Longrightarrow\quad\E'\cdot\E-\gamma\,\E^2=
\begin{cases}-\frac{\gamma^2}{\gamma+1}\left(\frac{\uel}{c}\cdot\E\right)^2&
\text{in Lorentz-group relativity,}\\
\hspace{11mm}0&\text{in groupoid relativity.}\end{cases}\end{gather}

Some other experimental predictions of the groupoid relativity, versus the predictions of the Lorentz relativity-group, are discussed in [Oziewicz 2006, 2007].

\section{Addition of ternary relative velocities}\label{ter}
\begin{table}[h!]\begin{center}\caption{Nonassociative $\oplus$-addition of reciprocal ternary relative velocities, no Figure. The addition of orthogonal relative velocities, $\vel\cdot\uel^{-1}=0,$ looks `the same' for binary and ternary relative velocities [Oziewicz 2005].\label{ass4}}
$\begin{array}{l}\\\hline\hline\\\vspace*{3mm}
P\cdot\uel=0,\quad P\cdot\vel=0,\quad\uel^{-1}=-\uel,\quad\vel^{-1}=-\vel
\qquad\Longrightarrow\\\vspace*{3mm}
\begin{aligned}\textstyle{\left(1-\frac{\vel\cdot\uel^{-1}}{c^2}\right)}\;
\vel\oplus_P\uel\quad&=\quad\textstyle{\uel+\frac{\vel}{\gamma_\uel}
-\frac{\gamma_\uel}{(\gamma_\uel+1)}\frac{(\vel\cdot\uel^{-1})}{c^2}
\;\uel}\\\vspace*{3mm}
&=\quad\uel\;+\;\vel\quad+\quad\textstyle{\frac{\gamma_\uel}{\gamma_\uel+1}}\,\uel^{-1}\cdot
\frac{(\vel\wedge\uel^{-1})}{c^2}\end{aligned}\\\hline\hline\end{array}$\end{center}\end{table}

Table \ref{ass4} shows the addition of relative velocities parameterizing the isometric Lorentz relativity-transformations, [Fock 1955, 1961 \S 16, formula (16.08), 1964].
\subsection{$\oplus$-inverse}\label{ter1} The $\oplus$-inverse is the reciprocal velocity, $\uel^{-1}=-\uel,$ as in the case of absolute time:
\begin{gather}\vel\oplus\uel=\zel\quad\Longleftrightarrow\quad
\vel+\uel=\zel,\notag\\\text{that is:}\quad
\oplus-\text{inverse}\;=\;(+)-\text{inverse}.\label{recip}
\end{gather}
\subsection{Mocanu paradox}\label{ter2} The coincidence of the Galilean $(+)$-inverse and the Einstein $\oplus$-inverse, gives the Mocanu paradox [Mocanu 1986]: $\oplus$-inverse is $\oplus$-automorphism,
\begin{gather}(\vel\oplus\uel)^{-1}=(\vel^{-1})\oplus(\uel^{-1})\quad\neq\quad
(\uel^{-1})\oplus(\vel^{-1}).\end{gather}  Whereas one would expect
that the unary inverse operation is an \textit{anti}-automorphism,
\begin{gather}(f\circ g)^{-1}=(g^{-1})\circ(f^{-1}).\end{gather}\vspace{-10mm}
\begin{figure}[h]\caption{Addition of ternary relative velocities: Mocanu paradox\label{Mocanu}}
$$\xymatrix{A\quad\ar@/^1pc/[r]^{\uel}\ar@/^3pc/[rr]^{\vel\oplus\uel}
&\quad B\quad\ar@/^1pc/[r]^{\vel}\ar@/^1pc/[l]_{-\uel}&\quad
C\quad\ar@/^1pc/[l]_{-\vel}\ar@/^2pc/[ll]^{(-\uel)\oplus(-\vel)\neq-\vel\oplus\uel}}$$
\end{figure}
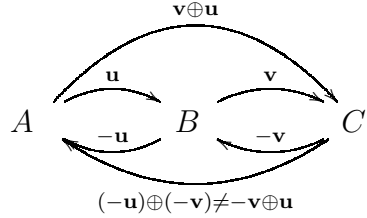\vspace{-10mm}

\subsection{Ungar's discovery: nonassociativity}\label{ter3} In this subsection we put $c^2=1.$ In 1988 Ungar discovered that the $\oplus$-addition is non-associative [Ungar 1988, p. 71],
\begin{gather}\begin{aligned}\frac{\gamma_{\wel\oplus(\vel\oplus\uel)}}{\gamma_\wel
\gamma_\vel\gamma_\uel}&=1+\uel\cdot\vel+\vel\cdot\wel+\wel\cdot\uel+
\frac{\gamma_\uel}{\gamma_\uel+1}(\wel\wedge\uel)\cdot(\uel\wedge\vel),\notag\\
\frac{\gamma_{(\wel\oplus\vel)\oplus\uel}}{\gamma_\wel\gamma_\vel\gamma_\uel}&=
1+\uel\cdot\vel+\vel\cdot\wel+\wel\cdot\uel+
\frac{\gamma_\uel}{\gamma_\uel+1}(\wel\wedge\vel)\cdot(\vel\wedge\uel),\end{aligned}\\
\{\wel\oplus(\vel\oplus\uel)\}\;\wedge\;\{
(\wel\oplus\vel)\oplus\uel\}=A(\wel\wedge\vel)+B(\vel\wedge\uel)+C(\uel
\wedge\wel)\;\neq\;0.\label{Ungar}\end{gather} Thus, not only are
these two resulting relative velocities, $\wel\oplus(\vel\oplus\uel),$ and $(\wel\oplus\vel)\oplus\uel,$ non collinear \eqref{Ungar},
but also their differ in their scalar \textit{magnitude}.

Many opponents disagree with non-associative law of addition of the Einstein relative velocities. Opponents claim that if a group $G$ is associative (think about composition of matrices), then the Lobachevski factor space $G/H$ must be necessarily an associative group, even if $H$ is not a normal subgroup, e.g. [Barrett 2006; Daniel Sudarsky, UNAM, ICN, 2003].

\section{Addition of binary relative velocities}
\begin{figure}[h!]\caption{Three body, $\{A,B,C\},$ in relative motions, $B\cdot\uel^{-1}=0$\label{vu}}
$$\xymatrix{\text{Alice}\quad\ar[r]^{\uel}\ar@/_2pc/[rr]^{\vel\circ\uel}
&\quad \text{Bob}\quad\ar[r]^{\vel}\ar@/_2pc/[l]_{\uel^{-1}}&\quad
\text{Clare}}$$\end{figure}
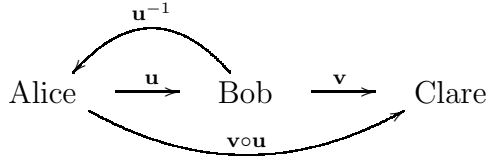

Consider a system of three bodies $\{A,B,C\},$ Figure \ref{vu}. Clare $C$ is moving with a binary velocity $\vel$ relative to Bob $B,$ and Bob $B$ is moving with a binary velocity $\uel$ relative to Alice $A.$ What is the binary velocity of Clare $C$ relative to Alice $A$?

\begin{table}[h]\begin{center}\caption{Associative $\circ$-addition
of binary relative velocities, Figure \ref{vu}. The addition of orthogonal
relative velocities, $\vel\cdot\uel^{-1}=0,$ looks `the same' for
binary and ternary relative velocities [Oziewicz 2005].\label{ass2}}
$\begin{array}{l}\\\hline\hline\\\vspace*{3mm}
A\cdot\uel=0,\quad B\cdot\vel=0,\quad B\cdot\uel^{-1}=0,\quad\uel^{-1}\neq-\uel,\quad\vel^{-1}\neq-\vel
\qquad\Longrightarrow\\\vspace*{3mm}
\begin{aligned}\textstyle{\left(1-\frac{\vel\cdot\uel^{-1}}{c^2}\right)}\;\vel\circ\uel\quad&=\quad
\displaystyle{\uel+\frac{\vel}{\gamma_\uel}
\quad+\quad\frac{\vel\cdot\uel^{-1}}{c}}A.\\\vspace*{3mm}
(\gamma_\uel^2-1)\textstyle{\left(1-\frac{\vel\cdot\uel^{-1}}{c^2}\right)}
\vel\circ\uel&=\left\{\gamma_\uel^2\textstyle{\left(1-
\frac{\vel\cdot\uel^{-1}}{c^2}\right)}-1\right\}\,\uel-
\gamma_\uel\,\uel^{-1}\cdot\frac{(\vel\wedge\uel^{-1})}{c^2}\end{aligned}
\\\hline\hline\end{array}$\end{center}\end{table}

\newpage
\section{Final remarks}
\begin{Group} An analysis of the derivation of the Lorentz group as the group of transformations relating observers, and the velocity $\oplus$-addition in [Einstein 1905], reveals that the inverse-velocity property \eqref{recip} was the most important tacit independent assumption used most effectively as an axiom [Einstein 1905 p. 901], and is not related to the verbal Einstein's two postulates [1905 pp. 891-892]. The reciprocal-velocity axiom \eqref{recip} tells that every observer measuring some velocity can measure also inverse of this velocity. It is however true that Einstein's reciprocal-velocity axiom \eqref{recip} is absolutely necessary for the derivation of the Lorentz \textit{group} as the one that relates two observers:
\begin{center}Lorentz group relating observers$\qquad\Longrightarrow\qquad$
\{$\oplus$-inverse\quad=\quad $(+)$-inverse\}.\end{center}\end{Group}

\begin{Thomas} The Thomas ro\-ta\-tion (Thomas in 1926, see [Ungar 1988]), means non-transitivity of the parallelism of the spatial frames. The addition of Einstein relative velocities, being non-associative, is not a group operation. It is a \textit{loop} operation [Ungar 1988]. Non-associative $\oplus$-addition is counterintuitive and paradoxical: for a system of four or more bodies the $\oplus$-addition of three non-collinear relative velocities gives the \textbf{two} distinct velocities between two bodies. There have been attempts [Ungar 1988] to explain the non-associativity, and also Mocanu paradox, as the Thomas ro\-ta\-tion, \ie as non-transitivity of the parallelism of the spatial frames. We consider this attempt not satisfactory. Jackson [1962] argued that the Thomas rotation is \textit{necessary} in order to explain factor `2' in the doublet separation for spin-orbit interaction. Einstein was surprised that Thomas's relativistic `correction' could give factor `2'. Dirac in 1928 explained the same factor and the correct spin levels in terms of the Clifford algebra and the Dirac equation, without invoking the Thomas rotation. The Dirac equation conceptually ought to be understood in terms of the Clifford algebra alone. No longer did anyone need Thomas's precession except for the non-associative $\oplus$-addition of velocities.\end{Thomas}

\begin{conclusion} Most readers will consider three properties \S \ref{ter1}-\ref{ter2}-\ref{ter3}, of the Lobachevsky manifold of Einstein's velocities, $O(1,3)/O(3),$ as being very attractive from the point of view of mathematics and physics. Moreover, these properties are consistent with the concept of the absolute preferred space, that explain the twin paradox [Herbert Dingle, Rodrigo de Abreu, Subhash Kak]. Does these attractiveness must forbid the consideration of the alternative theories of special relativity that does not need the absolute space?

Maybe some readers would like to consider the necessity of the absolute space to be a deficient property of such special relativity theory?
$$\begin{aligned}\text{No absolute space}\quad&\Rightarrow\quad\text{no Einstein's relative velocities}\\&\Rightarrow\quad\text{no Einstein's special relativity}\\& \Rightarrow\quad\text{no asymmetric biological ageing of twins}\end{aligned}$$
\end{conclusion}

\end{document}